\documentclass{article}

\usepackage{amsmath, graphicx, hyperref,  
}
\usepackage[shortcuts]{extdash}

\newcommand{\ket}[1]{| #1 \rangle}                     
\newcommand{\bra}[1]{\langle #1 \, |}                  
\def\pa{\partial}
\def\<{\langle}
\def\>{\rangle}

\def\tr{\mathrm{tr\, }}

\begin{document}


\title{Energy Positivity, Non-Renormalization, and Holomorphy in Lorentz-Violating Supersymmetric Theories}
\date{}
\author{ Adam B. Clark\footnote{aclark@muhlenberg.edu}\\
Department of Physics, Muhlenberg College\\
2400 Chew St., Allentown, PA 18104, USA} \maketitle

\begin{abstract}

This paper shows that the positive-energy and non-renormalization theorems
of traditional supersymmetry survive the addition of Lorentz violating
interactions. The Lorentz-violating coupling constants in theories using
the construction of Berger and Kostelecky must obey certain constraints in
order to preserve the positive energy theorem. Seiberg's holomorphic
arguments are used to prove that the superpotential remains
non-renormalized (perturbatively) in the presence of Lorentz-violating
interactions of the Berger-Kostelecky type.  We briefly comment on
Lorentz-violating theories of the type constructed by Nibbelink and
Pospelov to note that holomorphy arguments offer elegant proofs of many
non-renormalization results, some known by other arguments, some new.
\end{abstract}
\maketitle

\section{Introduction}
By employing the holomorphic arguments of Intriligator, Leigh, and Seiberg
\cite{Seiberg2}, one can show that the full non-renormalization theorems of
$\mathcal{N} =1$ supersymmetry apply unaltered to theories with Lorentz
violating (LV) interactions of either the Berger-Kostelecky (BK) type
\cite{Kostelecky2} or the Nibbelink-Pospelov (NP) type \cite{groot1}. The
essential point of the proof is that Lorentz symmetry plays no direct role
in the holomorphy argument. As long as the normal rules of $\mathcal{N}=1$
SUSY are followed when constructing the model, and as long as the LV
interaction creates no new anomalies or other surprises, then the
superpotential will be protected against perturbative quantum corrections,
and under appropriate conditions an exact expression for the quantum
effective superpotential can be obtained, using now-standard arguments from
\cite{Seiberg2}.

The rest of the paper is organized as follows:  first, we review the general
holomorphy arguments for non-renormalization in supersymmetric theories.
Next we examine BK-type theories, demonstrating that they satisfy the
conditions of Seiberg's holomorphy argument. Third, we show that BK-type
theories require additional constraints on the values of the LV coupling
constant in order for the positive energy theorem to hold. Next, we comment
briefly on NP-type theories, explaining how holomorphy arguments more or
less automatically prove that superpotential LV couplings and potentially
divergent FI terms are protected against perturbative corrections.
Holomorphy arguments go one step farther, and prove that the NSVZ
$\beta$-function (in holomorphic coupling) remains subject only to one-loop
renormalization and that NP-type LV couplings that enter into the
gauge-kinetic function are immune to perturbative renormalization (but still
subject to wavefunction renormalization). Finally, we summarize and
conclude.

\section{Non-Renormalization of Berger-Kosteleck\'y Models by Holomorphy}

\subsection{Review of Seiberg's Proof by Holomorphy in Standard Supersymmetric Theories}
\label{sec:review}

The arguments of Seiberg et al.\cite{Seiberg2} hinge on three key points: 1)
respect of symmetries, 2) holomorphy of the superpotential, and 3) the fact
that holomorphic functions are completely determined by their singularities
and asymptotic behavior \cite{Seiberg1}. All tree-level couplings in the
superpotential are treated as auxiliary fields, or fully-fledged chiral
superfields that just happen to be non-dynamical.  A coupling that
explicitly breaks a global symmetry of the rest of the theory in turn
provides a selection rule constraining quantum corrections:  since
symmetry-breaking terms in the quantum effective potential must ultimately
descend from tree-level breaking terms, we can employ the usual ``that which
is not forbidden is compulsory'' algorithm simply by pretending that the
coupling itself transforms in just the right way to preserve the broken
symmetry.  This provides a simple check on whether symmetry-breaking terms
in the effective superpotential are consistent with the tree-level breaking
terms.  This is how Seiberg's prescription respects all symmetries, even the
broken ones \cite{Seiberg1, Seiberg2}. Lorentz-violating theories themselves
almost invariably employ that technique for the LV couplings
\cite{Kostelecky1,Kostelecky2}.  In much of the Lorentz-violating
literature, these transformation properties of the LV couplings are dubbed
``observer Lorentz invariance.''  See \cite{donold,Kostelecky1} for detailed
discussions. In the recent work of \cite{kiritsis}, native to the AdS/CFT
correspondence, this phenomenon is referred to more simply as diffeomorphism
invariance.

Holomorphy of the superpotential is a proxy condition for invariance under
supersymmetry, given that one is constructing a theory using the formalism
of superfields.  In some sense, this is just another symmetry to respect,
but this symmetry is powerful enough to deserve special mention.
Supersymmetry is so restrictive that it enables divergence cancellations in
1-loop diagrams in the traditional, pre-Seiberg proofs of
non-renormalization.  Part of Seiberg's great insight was that
holomorphicity could be taken literally and was every bit as restrictive
mathematically as supersymmetry invariance is physically.  This leads to
point 3, which is the punchline:  Respect of symmetries makes it possible to
write down the most general \emph{holomorphic} function of couplings and
superfields for the superpotential.  Many coefficients are fixed outright by
the requirement of holomorphy.  Still more coefficients can be obtained by
analyzing the theory in some appropriate limit, since holomorphic functions
are completely determined by their singularities and their asymptotics.
Often these constraints will completely determine the superpotential
\cite{Seiberg1,Seiberg2}.

\subsection{Berger-Kostelecky Lorentz Violation}
\label{sec:proof}

Even spacetime symmetries could be viewed by a model-builder as ``just
another set of symmetries.''  In the BK-type theories, spacetime symmetries
are altered by broken Lorentz invariance, and the superalgebra is modified
\cite{Kostelecky2,patdon}.  In NP-type theories, spacetime symmetries are
altered, but the superalgebra is not \cite{groot1,groot2}. In both cases,
the theory can still be described in terms of superfields, and the
superpotential is a holomorphic function of said superfields. In the
BK-construction these superfields are not necessarily the same as the chiral
and vector superfields used in traditional SUSY.  Whether or not the
superalgebra is modified, invariance under the (possibly modified)
supersymmetry is still encoded in the holomorphy of the superpotential.

Berger and Kostelecky begin with an ordinary Wess-Zumino model, then add
Lorentz-violating interactions to the K\"ahler potential.\footnote{While
\cite{Kostelecky2} does not use this term explicitly, they do point out that
their LV interactions do not affect the superpotential.  This is not
obvious, since their construction involves modifying the superfields rather
than adding an LV interaction constructed out of superfields. We clarify
this point in section \ref{sec:proof}.} They then show that the resulting
Lagrangian is \emph{almost} invariant under ordinary supersymmetry but
becomes completely invariant (up to total derivative terms) under slightly
modified SUSY transformations \cite{Kostelecky2}. Fermion and boson
propagators are modified in the Lorentz-violating theories, but they retain
the parallel structure which is essential for brute-force proofs of
divergence cancellation in traditional SUSY theories, leading
\cite{Kostelecky2} to very plausibly assert that those divergences should
still cancel. Berger and Kostelecky construct modified chiral superfields
for their LV SUSY theories, which we will exploit to concisely prove that
Berger and Kostelecky were correct about the non-renormalization theorem and
divergence cancellation.

Berger and Kostelecky construct LV theories using Majorana spinors following
the notation conventions of Wess and Bagger's seminal work
\cite{WessBagger}. We begin by first rewriting in the slightly more modern
notation of \cite{Martin} and writing an LV Wess-Zumino model for a chiral
multiplet with Weyl spinors rather than Majorana.  Our chiral superfield for
normal SUSY theories is
\begin{align}
\Phi = \, & \phi(x) + i \theta^\dagger \bar{\sigma}^\mu \theta \pa_\mu \phi(x)
+\frac{1}{4} \theta \theta \theta^\dagger \theta^\dagger \pa_\mu \pa^\mu \phi(x)
 \\
\nonumber & + \sqrt{2} \theta \psi(x)- \frac{i}{\sqrt{2}} \theta \theta \theta^\dagger \bar{\sigma}^\mu \pa_\mu \psi(x)
+ \theta \theta F(x)
\end{align}
The usual Wess-Zumino Langragian in superfield form is given by
\begin{align}\label{eq:wzsuper}
\mathcal{L}_{WZ} & = \int d^4 \theta \ \Phi^* \Phi + \int d^2 \theta \ W(\Phi) + c.c.
\end{align}
with
\begin{equation}
W(\Phi) = \frac{M}{2} \Phi \Phi + \frac{g}{3} \Phi \Phi \Phi.
\end{equation}
More general theories could be constructed by promoting $W$ to an arbitrary
holomorphic function of $\Phi$ and replacing $\Phi^* \Phi$ with a more
general K\"ahler potential.  To facilitate contact with the work of Berger
and Kostelecky, we expand the basic Wess-Zumino lagrangian as
\begin{align}\label{eq:Lchiral}
\mathcal{L}_{WZ}  = & -\pa^\mu \phi^* \pa_\mu \phi + i \psi^\dagger \bar{\sigma}^\mu \pa_\mu \psi + F^* F \\
\nonumber  & + \left( - \frac{1}{2} M \psi^2 + M \phi F  - \frac{1}{2} g \phi \psi \psi \right) + c.c.
\end{align}

In the conventional picture (i.e. without using superfields), Lorentz-Violating
interactions are added in the form of the following term, $\mathcal{L}_{LV}$ \cite{Kostelecky2}
\begin{align}
\label{eq:llv}
\mathcal{L}_{LV} =& 2 k_{\mu\nu} \pa^\mu \phi^* \pa_\nu \phi + k_{\mu\nu} k^{\mu}_{\ \rho}
  \left( \pa^\nu \phi^* \pa^\rho \phi \right) \\
\nonumber  &
    + \frac{i}{2} k_{\mu\nu} \psi^\dagger \bar{\sigma}^\mu \pa^\nu \psi,
\end{align}
which can also be obtained from the original Lagrangian by replacing the
derivative operator with a so-called ``twisted'' derivative operator \cite{patdon}:
\begin{equation}
\label{twisted}
\tilde{\pa}_\mu = \left(\delta_{\mu}^{\ \alpha} + k_{\mu}^{\ \alpha}\right) \pa_\alpha.
\end{equation}
This operator is also denoted by $\nabla_m$ in \cite{petrov1}.  Indeed, many
quantities in conventional theories can be extended to BK theories by the
replacement $\pa_\mu \to \tilde{\pa}_\mu$ and ``twisting'' all vector
indices by the $\delta_{\mu}^{\ \alpha} + k_{\mu}^{\ \alpha}$ operator used
in \eqref{twisted} \cite{patdon}. This ``folk theorem'' extends to
superfields, as we see when looking at the LV version of the chiral
superfield \cite{Kostelecky2}:
\begin{align}
\Phi = & \phi(x) + i \theta^\dagger \bar{\sigma}^\mu \theta \left( \pa_\mu
+ k_{\mu\alpha} \pa^\alpha \right) \phi(x)  + \sqrt{2} \theta \psi(x) \\
\nonumber & +\frac{1}{4} \theta \theta \theta^\dagger \theta^\dagger
\left( \pa_\mu + k_{\mu\alpha} \pa^\alpha \right) \left( \pa^\mu
+ k^{\mu\beta} \pa_\beta \right) \phi(x) \\
\nonumber  & - \frac{i}{\sqrt{2}} \theta \theta \theta^\dagger \bar{\sigma}^\mu
\left( \pa_\mu + k_{\mu\alpha} \pa^\alpha \right) \psi(x) + \theta \theta F(x).
\end{align}

Building the LV interaction terms into a change of the superfield itself
obfuscates the nature of the LV interaction as belonging to the
superpotential or the K\"ahler potential. In \cite{Kostelecky2} it is noted
in passing that the LV interaction does not affect the superpotential.  To
understand this, note that the LV coupling $k_{\mu\nu}$ appears only in
terms including both $\theta$ and $\theta^\dagger$; therefore, since the
superpotential will only be integrated $\int d^2 \theta$ or $\int d^2
\theta^\dagger$, $k_{\mu\nu}$ will never appear in the action in a term born
of the superpotential. Thus, the LV interactions are best thought of as part
of the K\"ahler potential in the BK-construction.

When the full Lagrangian for the Lorentz-violating Wess-Zumino model with
one chiral multiplet is written by adding up the various pieces of the
Lagrangian (equations \eqref{eq:llv} and \eqref{eq:Lchiral}) in conventional
notation or by using the normal superfield Lagrangian \eqref{eq:wzsuper} but
with the modified LV superfields, the resulting theory is not quite
invariant under normal SUSY transformations \cite{Kostelecky2}.  If one
modifies the superalgebra and SUSY transformations by the same prescription
of ``twisting'' the derivative operator, then the modified Lagrangian is
invariant under the modified SUSY transformations \cite{Kostelecky2}.  In
summary, the Lagrangian $\mathcal{L} = \mathcal{L}_{WZ} + \mathcal{L}_{LV}$
is invariant under SUSY generators $Q$ and $Q^\dagger$ with superspace
representations
\begin{align}
Q & = i \pa_{\theta} - \sigma^\mu \theta^\dagger \pa_\mu - k_{\mu\nu} \sigma^\mu \theta^\dagger \pa^\nu \\
Q^\dagger & = i \pa_{\theta^\dagger} - \bar{\sigma}^\mu \theta \pa_\mu - k_{\mu\nu} \bar{\sigma}^\mu \theta \pa^\nu
\end{align}
and anti-commutation relation:
\begin{equation}
\label{eq:susytwisted}
\{Q,{Q^\dagger} \} =  2 \sigma^\mu \pa_\mu +  2 k_{\mu\nu} \sigma^\mu \pa^\nu,
\end{equation}
where $\sigma^0$ and $\bar{\sigma}^0$ are each the $2\times 2$ identity
matrix, $\sigma^i$ is the $i$th Pauli spin matrix, and $\bar{\sigma}^i = -
\sigma^i$.  We will strive to avoid the need for tracking spinor indices as
much as possible, but when unavoidable we follow \cite{Martin}.  In brief,
undotted Greek indices from the beginning of the alphabet ($\alpha, \beta,
...$) denote left-handed Weyl spinor indices while their dotted counterparts
denote right-handed Weyl spinor indices ($\dot{\alpha}, \dot{\beta}, ...$).
Spinor indices are implicitly raised and lowered as needed with the
two-index Levi-Civita $\varepsilon.$  Our only exception to leaving spinor
indices implicit is the gauge superfield strength, $W_\alpha$, which we
write out to distinguish from the superpotential, $W$.

There are some trivial but potentially confusing differences in notation.
Berger and Kostelecky use  $\theta$ and $\bar{\theta}$ where we use
$\theta^\dagger$ and $\theta$, respectively.  Invariance under the modified
SUSY transformations proceeds the same with Majorana or with Weyl spinors,
so we do not repeat the proof of invariance from \cite{Kostelecky2}.
Similar constructions exist for supersymmetric gauge theories, and we will
quote results from these theories only as needed.  The main difference
between the spinor conventions of \cite{Martin} and \cite{Kostelecky2} is
that the former removes the need for awkward-looking left- and right-handed
projection operators involving $\gamma^5$ by working with Weyl-spinors so
that undaggered spinors are implicitly left-handed and daggered spinors
right-handed.

The BK-construction for SUSY gauge theories is constructed similarly
\cite{patdon}. When writing out the vector superfield in terms of component
fields, simply ``twist'' each spacetime index on a field or derivative
operator with the $\left(\delta_{\mu}^{\ \alpha} + k_{\mu}^{\
\alpha}\right)$ operator.  Recasting the results of \cite{patdon} with
Weyl-spinors instead of Dirac we get
\begin{align}
V & = \theta^\dagger \bar{\sigma}^\mu \theta \left(\delta_{\mu}^{\ \nu} +
k_{\mu}^{\ \nu}\right) A_\nu + \theta^\dagger \theta^\dagger \theta \lambda
+ \frac{1}{2} \theta\theta \theta^\dagger \theta^\dagger D \\
W_\alpha & = -\frac{1}{4} D^\dagger D^\dagger D_\alpha V,
\end{align}
where the supercovariant derivatives are also twisted by the $\delta +k$
operator: $D_\alpha = \pa_{\theta^\alpha} - i (\sigma^\mu
\theta^\dagger)_\alpha \left(\delta_{\mu}^{\ \nu} + k_{\mu}^{\ \nu}\right)
\pa_\nu$.  The pure gauge Lagrangian is then the usual superspace integral
of $W_\alpha W^\alpha.$  This can be generalized to the non-Abelian case in
the usual way.  We emphasize that in this construction, the LV interactions
live entirely in the gauge-kinetic function, in contrast to the BK-model
with only chiral multiplets, where the LV interaction was implicitly part of
the K\"ahler potential.

\subsection{Non-Renormalization in Berger-Kostelecky Theories \label{BK}}
As discussed above, supersymmetric BK theories can be constructed out of
modified superfields with a superpotential which is an arbitrary holomorphic
function of those modified superfields\cite{Kostelecky2}, as with ordinary
SUSY. Holomorphy of the superpotential now encodes invariance under the
modified superalgebra.  Seiberg's holomorphy arguments
\cite{Seiberg1,Seiberg2} then apply in full, as they don't reference a
specific form of SUSY but more generally whatever (super)symmetry is
``proxied'' by holomorphy.  Non-supersymmetric Lorentz-violating theories
have been shown to be renormalizable in cases of pure gauge \cite{ymlv}, in
QCD \cite{qcdlv}, and in the electroweak sector \cite{ewlv}.  Additionally,
the renormalization of LV $\phi^4$ theory has been worked out to all orders,
and renormalization of LV Yukawa theories has been solved to one-loop order
\cite{altschul1}.  We conclude from this litany of examples that nothing
intrinsic to LV interactions impedes the standard program of
renormalization. Furthermore, BK-type LV interactions are not chiral in
nature and do not introduce any additional fermions, so they are not
expected to produce new anomalies.  We therefore conclude that the results
of \cite{Seiberg2} apply to supersymmetric BK theories.  It is worth noting
that  a brute force calculation using supergraphs has been carried out in
\cite{Redigolo} for BK theories with diagonal $k_{\mu\nu}$, confirming the
original suspicions of \cite{Kostelecky2} and proving non-renormalization in
the special case of diagonal $k_{\mu\nu}$.

Our holomorphy argument goes further and shows that all the
non\-/renormalization results of traditional SUSY apply to all
supersymmetric BK theories:  the superpotential is not renormalized at any
order in perturbation theory, although it may be subject to renormalization
through instantons or other non-perturbative effects. Additionally, such
non-perturbative renormalization can often be computed using the methods of
\cite{Seiberg2}. With Wess-Zumino models, such as studied in
\cite{Kostelecky2}, it is quite well known that Seiberg's arguments prove
the tree-level superpotential is \emph{exact}.  We have shown that this
continues in the presence of LV interactions, and this proof opens the door
to further Seiberg-style analysis of BK-type LV extensions to the MSSM.

The non-renormalization theorem goes beyond the LV Wess-Zumino model. Vector
superfields for BK-type theories were constructed in \cite{patdon}. As with
chiral superfields in \cite{Kostelecky2}, the prescription was to ``twist''
the derivative operator and all space-time indices.  Also as with chiral
superfields, the LV coupling appears only in terms with both $\theta$ and
$\theta^\dagger$, so the LV interaction is most properly thought of as part
of the K\"ahler potential.  The holomorphy argument is identical to the
chiral superfield case.  Furthermore, since practically any $\mathcal{N}=1$
SUSY theory can be built with a collection of vector and chiral superfields
with various interactions, our proof of non-renormalization for BK-type
theories extends quite broadly.  It is important to note, however, that the
LV coupling, as part of the K\"ahler potential in BK-type theories, is not
protected against renormalization.

\subsection{Robustness Against Coordinate Transformations}
A cautionary note has been pointed out numerous times \cite{groot1,patdon}
that the BK-type LV interactions can be absorbed into the metric by the
coordinate transformation ${x^\mu}' = x^\mu - k^\mu_{\ \nu} x^\nu$. It is
argued in \cite{patdon} that this coordinate transformation causes
Lorentz-violation to manifest itself in peculiarities of the coordinate
system, namely non-orthogonality. Nevertheless, BK-type LV interactions
could be realized outside the metric in a different setting. Any theory with
\emph{extended} SUSY and multiple sectors, one with BK-type LV and one
respecting Lorentz symmetry, would be immune to complete removal of the LV
interaction. In such a setup, the coordinate transformation to undo the LV
interaction in one sector would reintroduce it in the other sector. A simple
demonstration is $\mathcal{N}=2$ gauge theory with one hypermultiplet where
BK-type LV interactions exist for only one of the two $\mathcal{N}=1$ chiral
multiplets that comprise hypermultiplet.  The BK-type LV interaction would
partially break the SUSY down to $\mathcal{N}=1$, but attempting to undo the
LV interaction with a coordinate transformation would then swap the roles of
the two multiplets. Similar constructions have been  outlined for
$\mathcal{N}=1$ supersymmetry in \cite{patdon} and in \cite{Redigolo} where
the two sectors interact only via soft SUSY-breaking terms. Analogous
constructions could be used to partially break $\mathcal{N}=4$ to either
$\mathcal{N}=2$ or $\mathcal{N}=1$.

We emphasize that using BK-type LV interactions to partially break extended
SUSY results in a theory with manifest Lorentz violation.  Furthermore, the
details of both non-renormalization and energy positivity are largely
unchanged in the extended SUSY scenario.  Thus, we can view the original
$\mathcal{N}=1$ Berger-Kosteleck\'y construction as a laboratory for
exploring universal features of this class of Lorentz violating
supersymmetric theories.


We speculate that Seiberg's seminal results\cite{seibergN4}  for
$\mathcal{N}=2$ and $\mathcal{N}=4$ theories\footnote{The so-called analytic
``prepotential'' of $\mathcal{N}=2$ that determines all the dynamics is only
renormalized to one-loop order. The $\mathcal{N}=4$ theory is exactly
conformal.} will also continue to hold (in a sense) for BK-type theories.
These theories with Lorentz violation in extended SUSY were first
constructed in \cite{patdon}.  There are countless examples in the
literature of theories that break $\mathcal{N}=4 \to \mathcal{N}=2$,
$\mathcal{N}=4 \to \mathcal{N}=1$, or $\mathcal{N}=2 \to \mathcal{N}=1$
where the broken theory inherits many useful properties from the unbroken
theory, so partial breaking BK-type LV interactions should likewise inherit
many features from the unbroken theory. Our reasons are twofold: First,
analyticity/holomorphy is the centerpiece of Seiberg's arguments, and we
have shown that these arguments are unchanged by BK-type Lorentz violation.
Second, as discussed above, uniform BK-type Lorentz violation is equivalent
to a change of coordinates, and it does not seem credible that a change of
coordinates, however peculiar and non-orthogonal, could introduce running
couplings into a theory well known to be exactly conformal.  This would be
tantamount to an anomaly in the rescaling symmetry, which does not exist.

One might expect that an LV theory could develop unusual behavior rendering
the powerful methods of \cite{Seiberg2} inapplicable, but such concerns
prove groundless. For example, LV theories generically exhibit some form of
instability at Planck-scale energies.  Fortunately, these are reasonably
well understood in the LV literature and can usually be dealt simply by
taking the LV theory to be an effective theory with a UV-completion where
Lorentz symmetry is restored at some sub-Planckian scale \cite{Kostelecky1}.
As long as the cutoff scale for the effective theory is sufficiently below
the scale where instabilities develop, Lorentz symmetry is restored long
before any instability can develop, as has been thoroughly explained in
\cite{Kostelecky1}, for example. A second possibility is that modifying the
superalgebra will render it inconsistent.  For BK-type theories, this is not
the case, but care must be taken lest the energy positivity theorem be
destroyed.

\section{Energy positivity in the BK construction}
\label{sec:rest}

Examination of the modified superalgebra relation \eqref{eq:susytwisted}
reveals the concern at once. The operator $\{Q,Q^\dagger\}$ is positive
definite by construction.  In traditional SUSY this guarantees energy
positivity by well known arguments.  With the modified superalgebra of
BK-type Lorentz violation, positive definiteness of $\{Q,Q^\dagger\}$ can
actually require negative energy if the components of $k_{\mu\nu}$ are too
negative.  By inspection one can see that the choice $k_{00} < -1$, for
example, will require negative energy.\footnote{This was noted earlier in
\cite{petrov1}. The LV coupling in that case was restricted to the special
form $k_{\mu\nu} = \alpha u_\mu u_\nu$, where $u_\mu$ was a 4-vector of norm
$\pm 1$ or 0.} Clearly the components of $k_{\mu\nu}$ must be subject to
additional constraints if the positive energy theorem is to survive.

It is worth noting that ambiguities arise when defining the Hamiltonian for
the Dirac equation in the presence of Lorentz violation, and that it is
necessary to perform a spinor-field redefinition in order to have a
hermitian Hamiltonian for Dirac particles \cite{Kostelecky1}. Fortunately,
the redefinition of what is meant by the ``Hamiltonian'' and ``energy'' does
not impact this discussion, since the questions here relate to $p_{0}$, the
space-like $p_{i}$, and the LV coupling $k_{\mu\nu}$. The phrase ``energy
positivity'' describes the $p_0 \geq 0$ condition, and even after redefining
spinor fields, it remains true that $p_{0}$ is equal to the Hamiltonian.

In this section we take the expectation value of $\{Q, Q^\dagger \}$ for
various generic spin-0 and spin-1/2 states and explore the constraints on
$k_{\mu\nu}$ necessary to preserve the positive energy theorem.

\subsection{Constraint from spin 1/2 particles at rest}

Taking the expectation value of $\{Q, Q^\dagger \}$ for a generic spin 1/2
state, $\ket{\psi}$, yields the following modified positive energy
condition:
\begin{equation}
\label{susybase}
0 \leq \langle \psi | \left( \sigma^\mu (p_\mu + k_{\mu}^{\ \nu} p_\nu \right) | \psi \rangle
\end{equation}
We are interested in constraints on $k_{\mu\nu}$ such that \eqref{susybase}
guarantees $p_0 \geq 0$, i.e. energy positivity. We will evaluate this with
the assumption that $\ket{\psi}$ is a generic but normalized two-component
spinor, parametrized as $\ket{\psi} = \left(
\begin{matrix} a
\\ b \end{matrix} \right).$
This yields
\begin{equation}
\label{vectorcon}
\begin{split}
0 \leq \ &  (p_0 + k_0^{\, \alpha} p_\alpha) +  (p_3 + k_3^{\, \alpha}p_\alpha)\left( |a|^2 -|b|^2\right)  \\
    & +2(p_1 + k_1^{\, \alpha} p_\alpha) \mbox{Re}(a^* b)  + 2 (p_2 + k_2^{\,
    \alpha} p_\alpha) \mbox{Im}(a^* b).
\end{split}
\end{equation}
When evaluated in the rest frame of the particle, the inequality becomes
\begin{equation}
\label{fullcon}
\begin{split}
0 \leq \ &  p_0 \Bigl( 1 + k_0^{\, 0} + k_3^{\, 0}\left( |a|^2 -|b|^2 \right) \\
    & +2 k_1^{\, 0} \mbox{Re}(a^*b) + 2 k_2^{\, 0} \mbox{Im}(a^*b) \Bigr).
\end{split}
\end{equation}
This expression does not lend itself easily to analysis and completely
obscures the rotational symmetry of our theory (when $k_\mu^{\ \nu}$ is
taken to transform appropriately).  To simplify this expression, we note
that the terms of \eqref{fullcon} involving the $k_i^{\ 0}$ have the
structure of a dot product of two 3-vectors.  Define $\vec{k}= (k_{1}^{\ 0},
k_{2}^{\ 0}, k_{3}^{\ 0})$ and $\vec{a}=\left( 2\mbox{Re} (a^{*}b), 2
\mbox{Im} (a^{*}b), |a|^{2} - |b|^{2}\right)$. The vector $\vec{a}$ has unit
norm since spinor $\ket{\psi}$ is normalized. With this replacement,
equation(\ref{fullcon}) becomes manifestly invariant under rotations:
 \begin{equation}
 \label{restwithkvec}
0 \le p_{0} \left( 1+ k_{0}^{\ 0} +\vec{k} \cdot \vec{a} \right).
\end{equation}
We can now more easily explore different scenarios by considering the
orientation of the vector $\vec{a}$ relative to $\vec{k}$.  The case
$\vec{a}\perp\vec{k}$ gives us a constraint on $k_0^{\ 0}$ (also mentioned
above, obtained by inspection of \eqref{eq:susytwisted}):
\begin{equation}
\label{k0con}
k_0^{\, 0} > -1,
\end{equation}
where we have chosen strict inequality, since any value of $p_0$ would still
satisfy the inequality if we chose $k_0^{\ 0}=-1.$  Once we have fixed
$1+k_0^{\ 0}$ to be positive, the worst case scenario arises when $\vec{a}$
is chosen to be anti-parallel to $\vec{k}$.  Satisfying \eqref{restwithkvec}
with positive $p_0$ then requires
\begin{equation}
\label{nicecon}
|\vec{k}|= \sqrt{ \left(k_1^{\ 0} \right)^2  + \left(k_2^{\ 0} \right)^2
+ \left(k_3^{\ 0} \right)^2 } < 1
+k_{0}^{\ 0}.
\end{equation}
In other words, if $k_{0}^{\ 0}$ or $\vec{k}$ violate the bounds set by
\eqref{k0con} and \eqref{nicecon}, then there exists some spinor
$\ket{\psi}$ such that $p_{0}<0$ for that state in order to satisfy equation
(\ref{susybase}). Thus a BK theory violating either of those equations is
unstable in a manner that cannot be rectified with a UV completion.

Similar constraints were explored via the dispersion relation in
\cite{petrov1}, with the restriction to the case $k_{\mu\nu} = \alpha u_\mu
u_\nu$.  They found that $|\alpha| \ll 1$ together with $u_\mu u^\mu = \pm
1,\ 0$ were sufficient to ensure consistency and that the LV terms could be
treated as ``small corrections.''  We go beyond the ``small correction''
case here to explore more detailed constraints for future model builders that may
succeed in finding additional SUSY-scale suppression of LV coupling
constants.

\subsection{Constraints from scalar particles}
Let us now evaluate \eqref{susybase} with scalar states instead of fermions.
The equation becomes
\begin{equation}
\label{scalarbase}
\begin{array}{lcl}
0  & \leq  & \bra{\phi} \{Q,\bar{Q}\} \ket{\phi} = \bra{\phi} \left( 2 \sum_{\mu=0}^3
(p_\mu + k_{\mu}^{\ \nu}
p_\nu )
\right) \ket{\phi} \\
    &= & \bra{\phi} 2 \Biggl[ p_0 \left(1 + \sum_{\mu=0}^3 k_\mu^{\ 0}\right) + p_1
    \left(1 + \sum_{\mu=0}^3
    k_\mu^{\ 1}\right) \\
    & & + p_2 \left(1 + \sum_{\mu=0}^3 k_\mu^{\ 2}\right)+ p_3 \left(1 +
    \sum_{\mu=0}^3 k_\mu^{\ 3}\right)
    \Biggr] \ket{\phi}.
\end{array}
\end{equation}

A simple starting expression is obtained by evaluating this in the rest frame of the state $\phi$, we see that $p_0 \geq
0$ is guaranteed only if
\begin{equation}
\label{scalarcon}
\sum_{\mu=0}^3 k_\mu^{\ 0}  \geq -1.
\end{equation}
If the $\mu 0$ components of $k$ violate this inequality, then any state
with scalar particles necessarily has negative energy, even when the
particles are at rest.

Equation \eqref{scalarbase} can be used to obtain more general constraints
by plugging boosted values of 4-momentum.  This is discussed below in
section \ref{moving}.

As mentioned earlier, stringent phenomenological limits on the size of
Lorentz violating couplings exist.  For the non-supersymmetric Standard
Model Extension, the results of recent literature are nicely summarized and
tabulated in \cite{data}. The supersymmetric LV parameter $k_{\mu\nu}$ is
related to the non-SUSY $c$ and $k_F$ coefficients from \cite{data}.  The
most forgiving of these constraints is $O(10^{-10})$, so consistency
constraints \eqref{k0con} and \eqref{nicecon} are more or less automatically
satisfied in any phenomenologically interesting theory.  However, should a
means be found to give Berger-Kostelecky twisted SUSY-LV couplings
additional suppression of order the SUSY-breaking scale (as has been done
with non-twisted SUSY-LV in \cite{groot1}), such a theory would need to
respect these $O(1)$ constraints.

\subsection{Constraints from moving particles}
\label{moving}
\subsubsection{General form of constraint}
If we allow any $p_i$ to be non-zero, then the bound of \eqref{nicecon} no
longer applies. We must re-examine the constraint condition
\eqref{vectorcon}. We first consider, for simplicity, a particle moving in
the 1-direction.  Instead of \eqref{fullcon}, we now find
\begin{equation}
\label{boostedbase}
\begin{split}
0 \leq & p_0 \Bigl( 1 + (k_0^{\, 0}) + (k_3^{\, 0})\left( |a|^2 -|b|^2 \right)
    +2 (k_1^{\, 0}) \mbox{Re}(a^*b) \\
    & + 2 (k_2^{\, 0})
    \mbox{Im}(a^*b) \Bigr) + p_1 \Bigl((k_0^{\, 1}) + (k_3^{\, 1})\left( |a|^2 -|b|^2 \right) \\
    & +2 (1+k_1^{\, 1}) \mbox{Re}(a^*b) + 2
    (k_2^{\, 1}) \mbox{Im}(a^*b)
    \Bigr).
\end{split}
\end{equation}
This can be simplified by use of the previously introduced vector $\vec{k}$
and a new vector that captures information about the space-space components
of the second row of $(k_\mu^{\ \nu})$.  Let
\begin{equation}
\label{R}
\vec{R} = \left(1+  k_1^{\ 1}, k_2^{\ 1}, k_3^{\ 1}
\right).
\end{equation}
Then, mirroring the procedure that led to \eqref{restwithkvec} we can
reorganize \eqref{boostedbase} as
\begin{equation}
\label{boostedvec}
0 \leq p_0 \left( 1 + (k_0^{\ 0}) + \vec{k} \cdot \vec{a}
    \right) +
    p_1 \left( (k_0^1) + \vec{R} \cdot \vec{a}
        \right).
\end{equation}
This makes it easy to generalize to the case of arbitrary 3-momentum by
introducing one such $\vec{R}$ for each space direction. Define a more
general construction of $\vec{R}^i$ as
\begin{equation}
\label{Rcomps}
\left( \vec{R}^i\right)_j =  \delta_j^{\ i} + k_j^{\ i}.
\end{equation}
Then the general SUSY constraint equation for arbitrary 3-momentum is
\begin{equation}
\label{3momcon}
\begin{split}
0 \leq &\ p_0\left(1 + k_0^{\ 0} +\vec{k}\cdot\vec{a}
    \right) + p_1 \left(k_0^{\ 1} + \vec{R}^1 \cdot \vec{a}
    \right) \\
    & + p_2 \left( k_0^{\ 2} + \vec{R}^2 \cdot \vec{a} \right) + p_3 \left(k_0^{\ 3} + \vec{R}^3 \cdot \vec{a}
    \right),
\end{split}
\end{equation}
It will simplify computations to rearrange this expression into a term which
is constant for all choices of spinor and a dot product term which varies
from spinor to spinor as follows:
\begin{equation}
\label{eq:pencons}
0 \leq  \left[ p_0 (1 + k_0^{\ 0}) + \sum_i p_i k_0^{\ i} \right] + \left[ p_0 \vec{k}
    + \sum_i p_i \vec{R}^i \right] \cdot \vec{a}.
\end{equation}
The worst case scenario occurs when $( p_0 \vec{k} + \sum_i p_i \vec{R}^i)$
is anti-parallel to $\vec{a}$, so the strictest constraints from
\eqref{eq:pencons} are
\begin{equation}
\label{eq:bestcon}
0 \leq   p_0 (1 + k_0^{\ 0}) + \sum_i p_i k_0^{\ i}  - \left|  p_0 \vec{k}
    + \sum_i p_i \vec{R}^i  \right|.
\end{equation}

Similarly, equation \eqref{scalarbase} now applies in full when we consider moving scalar particles.

There are two ways to further think about constraints \eqref{scalarbase} and \eqref{eq:bestcon}:  First, we can
obtain the momentum by applying a particle boost (i.e. a boost that does not
affect $k_{\mu\nu}$); second, we can impose a mass-shell condition on the
momentum.

\subsubsection{Boosted particles}
Consider first a boost in the 1-direction, such that $p'_0 = \gamma p_0$ and
$p'_1 = - v \gamma$, where $\gamma = 1/\sqrt{1-v^2}$ as usual.  Under this
boost, equation \eqref{eq:bestcon} becomes
\begin{equation}
\label{eq:boostcon}
\begin{array}{rl}
0 \leq & \gamma p_0 (1 + k_0^{\ 0}) + -v \gamma p_0 k_1^{\ 0}  - \left|  \gamma p_0
\vec{k}
    + -v \gamma p_0 \vec{R}^1  \right| \\
=& \gamma p_0 \left(1 + k_0^{\ 0} - v  k_1^{\ 0} - \left|\vec{k} - v \vec{R}^1 \right| \right) \\
= & \gamma p_0 \Biggl(1 + k_0^{\ 0} - v  k_1^{\ 0}  - \Bigl[\left( k_1^{\ 0}- v(1+k_1^{\ 1})  \right)^2  \\
    & +\left(k_2^{\ 0} - v k_2^{\ 1} \right)^2 +\left(k_3^{\ 0} - v k_3^{\ 1} \right)^2 \Bigr]^{1/2} \Biggr).
\end{array}
\end{equation}
The generalization to arbitrary boosts is straightforward but
unilluminating.  Consistency of the twisted superalgebra then demands the
components of $k_{\mu\nu}$ are chosen so that no choice of boost speed $v$
violates inequality \eqref{eq:boostcon} and its generalizations unless $v$
is high enough that the UV completion of the LV theory should be used, i.e.
if $\gamma p_0$ is greater than the cutoff scale.  We find it convenient to think about this in the following way:
$k_{00}$ sets a scale for the upper limit of the absolute value of the other
components of $k_{\mu\nu}$

\subsubsection{Enforcing the Mass Shell or Dispersion Relation}
Another important feature of BK-type LV theories is the modification of the
dispersion relation of particles due to Lorentz violation
\cite{Kostelecky1} . Instead of $p^2 = -m^2,$ the appropriate relation is
\begin{equation}
\label{eq:dispersion}
\left( (g^{\mu\nu} + k^{\mu\nu}) p_\nu\right)^2 = -m^2,
\end{equation}
as found by looking at the propagator of the fields in \cite{Kostelecky1}.
Lorentz violating terms in the Lagrangian change the pole of the propagator
in precisely the same way as simply applying the ``twisting'' rule of thumb
to the traditional relation.  We may profitably think about this as a
modification to the mass shell condition.  Where we normally think of the
on-shell condition as $-m^2 = -p_0^2 + \vec{p}^2$ (with $\vec{p}$ denoting
the space-like components of momentum), the relationship is now much more
complicated, with direction-dependent corrections to the old terms (arising
from the diagonal elements of $k_{\mu\nu}$) and the addition of cross-terms (arising from off-diagonal
elements of $k_{\mu\nu}$).

Consider, for simplicity, a particle moving purely in the 1-direction.  In a
Lorentz invariant theory the mass shell condition would require $p_1 =
\pm \sqrt{p_0^2 - m^2}$.  In a BK-type LV theory, that relation becomes
\begin{equation}
\label{onshell1}
\begin{array}{rl}
p_1 = & \frac{p_0}{2 (1 + 2 k^{11} + (k^2)^{11})} \Biggl( -(4 k^{01} + 2 (k^2)^{01}) \\
    & \pm \Bigl((4 k^{01}
+ 2 (k^2)^{01})^2  + 4 (1 + 2 k^{11} + (k^2)^{11}) \times \\
    & \left((1 + 2 k^{00} + (k^2)^{00})
- m^2 /p_0^2  \right) \Bigr)^{1/2}  \Biggr),
\end{array}
\end{equation}
where $(k^2)^{\mu\nu} = k^{\mu\alpha}k_\alpha^{\ \nu}$.  Note that similar
but less general constraints from the dispersion relation have been obtained
in \cite{petrov1} for the form of $k_{\mu\nu}$ considered there.

For a particle moving in a fixed direction, the mass shell condition
can be used as a constraint to eliminate one of the space-like components of
momentum in favor of an expression similar to \eqref{onshell1}.

A general boost would be parameterized by the three components of boost
velocity, $v_i$, subject to the constraint $\sqrt{v_1^2 + v_2^2 +v_3^2} \leq
1$. The mass shell condition could be used to eliminate one of these degrees
of freedom in favor of a constraint of the form of \eqref{onshell1}.  The
resulting expression is complicated and unilluminating.  A better use of this constraint in model-building would be to first propose a
choice of $k_{\mu\nu}$ then check to see whether \eqref{eq:bestcon} and
\eqref{scalarbase} can be violated for some on-shell choice of momentum.

\subsection{An Alternate View on the Positive Energy Constraints}
The results from section \ref{sec:proof} on the non-renormalization theorem
had a reassuring interpretation when viewed from the complementary
perspective of the transformation that ``undoes'' the LV interaction in a single-sector non-extended SUSY theory:
${x^\mu}' = x^\mu - k^\mu_{\ \nu} x^\nu$.  It would be very disturbing
indeed if a simple linear coordinate transformation invalidated the
non-renormalization theorem or the positive energy theorem, unless the
coordinate transformation was singular or otherwise illegal.  An obviously illegal choice of  $k_{00} = -1$ marks a theory
that transparently violates SUSY's positive energy theorem.  Viewed as a
coordinate transformation, it is equally obvious that the transformation is
singular if any diagonal element of $k$ equals $-1$. However, $k_{00} < -1$
continues to violate the positive energy theorem, whereas the coordinate
transformation is no longer singular, but would change the signature of the
metric.  A natural first guess is that legal choices of $k_{\mu\nu}$ correspond to coordinate
transformations that preserve the signature of the metric, or even the signs of
all the diagonal entries.  Enforcing this condition requires, for example,
\begin{equation}
-2 k_{00} + k_0^{\ \mu} k_{\mu 0} >1,
\end{equation}
which is not obviously related to the other constraints on energy positivity
from this section.  We conjecture that some appropriate
condition exists for $k_{\mu\nu}$ when viewed as a coordinate
transformation that captures both the rest frame constraints as well as the
boosted particle constraints.

\section{Non-Renormalization of Nibbelink-Pospelov Type LV Theories}
\subsection{Review of Nibbelink-Pospelov Construction}
The approach of Nibbelink and Pospelov (NP) does not alter the superalgebra.
Rather, they construct LV operators that explicitly break the boost part of
the superalgebra but preserve the subalgebra generated by translations and
supercharges only \cite{groot1}.  Their construction is native to superspace
and follows the usual convention of a holomorphic superpotential and
non-holomorphic K\"ahler potential.  As with the BK construction, Nibbelink
and Pospelov work with 4-component Dirac spinors in the language of Wess \&
Bagger \cite{WessBagger}.  We apply the same translations to the conventions
of \cite{Martin} as we did with the BK-construction.

Nibbelink and Pospelov classify the possible types of LV operators
consistent with exact SUSY up to dimension 5.  We list here for reference
those LV operators relevant for SUSY gauge theories.
Charged chiral superfields have only a single K\"ahler potential term at
dimension 5 (and none at lower dimensions) \cite{groot1}:
\begin{equation}
\label{kahlerscalar}
N^\mu \bar{\Phi} e^V \mathcal{D}_\mu \Phi.
\end{equation}
The gauge sector has one dimension 4 K\"ahler term \cite{groot1}:
\begin{equation}
\label{kahlervector}
a^{\dot{\alpha} \alpha} \tr \bar{W}_{\dot{\alpha}} e^V W_\alpha e^{-V},
\end{equation}
and three superpotential or gauge-kinetic terms \cite{groot1}:
\begin{equation}
\label{eq:weird}
 b^{\alpha \beta} \tr W_{(\alpha} W_{\beta)}, \ \
 c^{ \alpha \beta} \tr \Phi W_{(\alpha} W_{\beta)} ,\ \
 T^\lambda_{\mu\nu} \tr W_\alpha \sigma^{\mu\nu}_{\alpha\beta} \pa_\lambda W_\beta,
\end{equation}
where parentheses denote symmetrization of indices, and the gauge super
field strength $W_\alpha$ is given by
\begin{equation}
W_\alpha = - \frac{1}{4} D^\dagger D^\dagger \left( e^{-V} D_\alpha e^V \right).
\end{equation}
We take pains to distinguish the holomorphic superpotential $W$ from the
gauge superfield-strength $W_\alpha$ by always including the spinor index of
the latter, even when contracted.  To accommodate this convention, we have
used the notation of \cite{groot2} for this operator but made the spinor
indices explicit.

The operators of \eqref{eq:weird} all represent modifications to the
gauge-kinetic function.  As \cite{groot1} explains, only the last term of
\eqref{eq:weird} is non-vanishing for SQED or SQCD, and even that is only
true for SQED. The gauge super field strength $W_\alpha$ is gauge invariant
only for a U(1) group, but replacing the ordinary spacetime derivative with
a covariant derivative destroys the chirality condition, making the term not
supersymmetric.

\subsection{Non-renormalization in NP-type theories \label{sec:NPnonren}}
As in the BK-construction, holomorphy is key.  With the NP-construction, the
superalgebra is unmodified, so holomorphy of the superpotential encodes
invariance under traditional SUSY.  Thus, even with NP-type LV interactions,
the superpotential is immune to perturbative renormalization, and even
non-perturbative renormalization is subject to tight controls.  None of the
well-known LV operators in the NP-construction can be added to the
superpotential of SQCD, so we do not exhibit an exact superpotential
calculation.  An SQCD model including NP-type LV interactions in the gauge
superpotential or in the K\"ahler potential would at most alter the running
of the gauge coupling, changing Seiberg's results only by altering the
coefficient of the beta function.

Weinberg \cite{weinberg1} extends Seiberg's proof in three important ways:
First, he extends the SUSY non-renormalization theorems to
\emph{non-renormalizable} theories, so one need not worry that higher
dimension LV operators ruin these familiar results.  Second, he clearly
demonstrates that superpotential terms dependent on the gauge superfield
strength $W_\alpha$ are also protected against perturbative renormalization.
Third, he proves that FI terms in U(1) theories are also non-renormalized,
as long as the U(1) charges of all the chiral superfields add to zero.  This
condition is already a well-known necessity for anomaly cancellation, and is
included in the SQED model considered by \cite{groot1} as well as the richer
models of \cite{groot2}.

Weinberg's argument about the FI term has to do with gauge invariance. After
promoting the FI coupling to a superfield, the FI term would not be gauge
invariant if the coupling depended on any other superfield.  The only
gauge-invariant correction to the FI coupling arises from a diagram that
vanishes when the charges are chosen as above \cite{weinberg1}.

These conditions do not change in the presence of NP-type Lorentz violation.
This provides a very elegant alternative proof of the result from \cite{groot2} that NP-type
LV interactions do not induce a potentially divergent FI term.  If the FI
term is not present in the bare Lagrangian, it will not be induced in the
effective Lagrangian, even by LV interactions. Conversely, if the bare
Lagrangian includes an LV coupling in an FI term, that coupling will be
protected against perturbative renormalization.  This can be seen simply via
holomorphy, without the need for computing divergent loop diagrams.

Additionally, SUSY gauge theories are subject to powerful restrictions on
the renormalization of the gauge coupling.  When the fields are normalized
with ``holomorphic'' coupling, the famous NSVZ $\beta$-function is only
renormalized at one-loop order.  When the fields are rescaled to canonical
normalization, the $\beta$-function has a slightly different form but is
\emph{exact}.  The original results were obtained in \cite{nsvz,Shifman}. An
alternative derivation of the same results was obtained in
\cite{ArkaniHamed}.  An illuminating discussion of the alternative
computation can be found in \cite{Strassler}. However, Weinberg offers an
interesting proof of the one-loop-only renormalization result that holds for
arbitrary superpotential interactions and arbitrary gauge-kinetic function
couplings \cite{weinberg1}.  We briefly summarize his technique here and
extend it to Lorentz-violating theories.

Weinberg begins by using Seiberg's spurion prescription, treating new
coupling constants as background superfields with appropriate transformation
properties for maintaining all symmetries of the lagrangian.  Of particular
importance will be the R-charge of the coupling and the nature of the
coupling as either a chiral or vector superfield.  New interactions in the
K\"ahler potential must have vector superfield couplings, and as such these
new coupling constants can only appear in non-perturbative corrections to
the chiral pieces of the action, namely the superpotential and the
gauge-kinetic function.  So the K\"ahler LV interactions of
\eqref{kahlerscalar} and \eqref{kahlervector} cannot contribute to
perturbative renormalization of the effective superpotential or
gauge-kinetic function. Weinberg then counts the number of graphs of
different types that could contribute to a term in the effective
superpotential and/or gauge-kinetic function. He considers graphs with $E_V$
external gaugino lines and an arbitrary number of external $\Phi$ lines,
$\Phi$ being any component field of the chiral superfield(s) charged under
the gauge group, and $I_V$ internal V-lines, V denoting any component of the
vector superfield.  Let $\mathcal{A}_m$ denote the number of pure gauge
vertices with $m \geq 3$ V-lines, which will bring factors of the
holomorphic coupling, $\tau$.  Let $\mathcal{B}_{mr}$ denote the number of
vertices with $m\geq r$ V-lines and any number of $\Phi$ lines which arise
from extra terms in the gauge-kinetic function with $r$ factors of the
vector superfield strength, $W_\alpha$.  In the tree-level Lagrangian, the
coefficient of such interactions is denoted $f_r$, so each of these diagrams
will bring a factor of the appropriate $f_r$.  Finally, let $\mathcal{C}_m$
denote the number of vertices with 2 $\Phi$ lines and $m \geq 1$ V-lines,
arising from the traditional K\"ahler potential term, $\Phi^\dagger e^{-V}
\Phi$.  Matching gauge lines with the various types of vertices yields
Weinberg's relation:
\begin{equation}
2 I_V + E_V = \sum_{m \geq 3} m \mathcal{A}_m + \sum_r \sum_{m \geq r} m
\mathcal{B}_{mr} + \sum_{m \geq 1} m \mathcal{C}_m.
\end{equation}
Diagrams corresponding to the $\mathcal{A}_m$ and $\mathcal{C}_m$ terms
arise from standard terms in SUSY gauge theories. Those corresponding to
$\mathcal{B}_{mr}$ terms come from new interactions in the gauge kinetic
function. Since $W_\alpha$ has R-charge +1, the couplings $f_r$ must have
R-charge $2-r$ for the new term to have the requisite R-charge of +2 for
gauge-kinetic terms. Since the gaugino component of the vector superfield
has R-charge +1, the focus on graphs with $E_V$ external gauginos enforces a
relationship between $E_V$ and the coefficients $\mathcal{B}_{mr}$ appearing
in the diagram, which in turn allows one to compute the number of factors of
the gauge coupling in terms of the $\mathcal{A, B, C}$ coefficients.
Enumerating the possibilities shows that only five distinct choices of the
$\mathcal{A, B, C}$ coefficients are legal, all of which contribute graphs
independent of the gauge coupling, and all of which allow only a single
non-zero coefficient.  One of the five choices allows for a single
$\mathcal{B}_{mr}=1$, which is just the tree-level contribution; the other
choices only turn on an $\mathcal{A}$ or $\mathcal{C}$.  Exhausting this
enumeration shows that the gauge coupling only receives perturbative
corrections at the one-loop level and that all coefficients $f_r$ in the
gauge-kinetic function receive no perturbative corrections, apart from
wavefunction renormalization which is not addressed by Weinberg's argument.

Thus all LV coupling constants for superpotential or gauge-kinetic function
terms in the NP-construction are protected against perturbative
renormalization.  In \cite{groot2}, $\beta-$functions were computed to
first-order in LV for all LV couplings relevant to $\mathcal{N}=1$ SQED. Our
results are in perfect agreement with their findings for the beta function
$T^\lambda_{\mu\nu}$ from \eqref{eq:weird}.  We take this one step further,
showing that to \emph{any }order in the LV couplings, this term is only
subject to wavefunction renormalization.

\subsection{Berger-Kosteleck\'y Models with Charged Matter}
Now that we have fleshed out Weinberg's argument, we can apply it also to
the BK-construction for pure gauge theories.  Since the LV coupling enters
into redefinitions of the vector superfield, it will be part of the
gauge-kinetic function, and thus protected against perturbative corrections.
It is an interesting puzzle whether the BK-construction can accommodate
charged matter, as the gauge- and matter-sector LV couplings appear to have
such wildly different renormalization properties.  It may simply not be
possible. Another, more tantalizing possibility is that quantum effects
might force the LV couplings to differ in the two sectors, thereby breaking
supersymmetry. A third possibility is that gauge-chiral interactions will
cancel against pure chiral interactions and ultimately protect the LV
coupling, $k_{\mu\nu}$, despite its presence in the K\"ahler potential.

\subsection{A Comment on the Possibility of SUSY-scale Suppression of LV Couplings}
There is some discrepancy in the literature over whether SUSY breaking
effects can lead to additional suppression of Lorentz violating couplings.
When Lorentz violation in a Wess-Zumino model occurs via the cutoff
regularization procedure as studied in \cite{ralston1}, it is found quite
generally that quantum effects rescale Lorentz violating couplings by a term
proportional to $(M/\Lambda)^2 \log(M/\Lambda)$, where $M$ is the SUSY scale
and $\Lambda$ is the Lorentz-breaking scale.  On the other hand, the results
of \cite{Redigolo} indicate that SUSY-scale suppression of LV couplings is
incompatible with gauge theories and can only occur with neutral chiral
superfields.

While each of these works looks at different models of Lorentz violation,
the ``no-go'' results of \cite{Redigolo} for LV SUSY gauge theories are
compatible with our results.  Generically, we find that LV interactions in
the superpotential or the gauge-kinetic function are protected against
perturbative renormalization by an extension of Seiberg's holomorphy
arguments.  Those concerned with fine-tuning problems will need to consider
more exotic models than the original BK- and NP-constructions.

\section{Conclusion}
Lorentz symmetry is not a necessary ingredient in Seiberg's holomorphy
arguments.  Thus, Lorentz-violating SUSY theories of both Berger-Kostelecky
and Nibbelink-Pospelov type preserve all the divergence-cancellation and
non-renormalization aspects of traditional SUSY theories.

NP-type theories always preserve SUSY's positive energy theorem since they
do not alter the superalgebra, and LV couplings in superpotential terms are
protected against perturbative renormalization.  While the LV couplings are
still subject to non-perturbative effects, this is limited to wave-function
renormalization, and Seiberg's techniques for obtaining exact quantum
superpotentials continue to apply.  K\"ahler potential LV interactions are
not protected.   K\"ahler potential as well as gauge field-strength
superpotential LV interaction terms may alter the gauge-coupling beta
function and in turn change some of the constants in the exponents of
Seiberg's exact formulas, but in the absence of matter LV terms in the
superpotential, Seiberg's exact results are altered only in a trivial way
\cite{Seiberg2}. The NP construction of LV superpotential terms does not
appear compatible with gauge invariance for any but abelian gauge theories,
so LV terms that might have a more dramatic impact on Seiberg's results are
disallowed \cite{groot1,groot2}.

In BK-type theories, the single LV interaction is built into a redefinition
of the superfields and the superalgebra itself.  The construction is such
that the LV coupling constant will only survive Grassman integration in the
K\"ahler potential, so the superpotential remains unchanged.  Seiberg's
holomorphy arguments guarantee that the superpotential in BK-type theories
remains non-renormalized (perturbatively), but this offers no protection to
the LV coupling constant itself in Wess-Zumino models.

The positive energy theorem only continues to hold if the LV coupling,
$k_{\mu\nu}$ obeys constraints \eqref{k0con}, \eqref{nicecon},
\eqref{scalarcon}, in addition to \eqref{scalarbase} and \eqref{eq:bestcon},
which must hold for arbitrary on-shell momentum below the cutoff scale of
the effective theory.  While these constraints are many orders of magnitude
less stringent than current phenomenological limits, they become important
in models with O(1) LV couplings that are suppressed as we run to lower
energies. As noted above, such suppression requires different LV couplings
for gauge and matter multiplets, which typically requires some level of SUSY
breaking itself.

We have laid out such an example whereby BK-type LV interactions can be used
to partially break extended SUSY, rendering a theory that possesses both
(some) unbroken SUSY and BK-type LV interactions that are robust against
coordinate transformations.  We currently know of no such model that has
been fully fleshed out in the literature.  We save detailed investigation of
such models for the future, noting for the time being that the positive
energy theorem has essentially the same form regardless of the degree of
supersymmetry.

This work opens the door to the application of powerful modern techniques in
supersymmetry, such as Seiberg's holomorphy arguments, to theories with
Lorentz-violation. To our knowledge, the main body of the Lorentz violation
literature has not yet employed these techniques.\footnote{Two recent works
\cite{sundrum, kiritsis} approach Lorentz violation in a context native to
the AdS/CFT correspondence and quite different at first appearance from the
well-trodden paths initially blazed by Kostelecky and collaborators.  While
SUSY is an important ingredient in any AdS/CFT construction, the absence of
superfields in $\mathcal{N}=4$ SUSY necessarily renders these approaches
quite different from both the BK- and the NP-constructions.} It would be
interesting to explore the ``mixed sector'' approach to BK-type
Lorentz-violation mentioned above as well as NP-type theories with extended
supersymmetry, in particular $\mathcal{N}=4$ to compare with the general
AdS/CFT computations of \cite{kiritsis}.  It will also be particularly
interesting to consider BK-type Lorentz violation in the context of
$\mathcal{N}=2$ gauge theory with matter, where the Lorentz violation
affects only the matter multiplets.  The machinery of Seiberg-Witten theory
should apply, with SUSY breaking originating from the LV couplings rather
than mass terms.

\section*{Acknowledgements}
The author would like to thank A.~Karch, A.~Nelson, D.~P.~Wilson,
P.~McDonald, D.~Colladay, M.~Huber, W.~Gryc, and B.~Fadem for helpful
conversations and comments on earlier drafts.  The author gratefully
acknowledges the generous support of the Office of the Provost of Muhlenberg
College via a Faculty Summer Research Grant.

\end{document}